\documentclass[10pt,twocolumn]{article}
\usepackage{graphicx} 
\graphicspath{{figures}} 
\usepackage[font=small,labelfont=bf]{caption} 
\usepackage{makecell}
\usepackage{svg}
\usepackage{textcomp}
\usepackage{stfloats}

\usepackage{xcolor}
\definecolor{black}{HTML}{212427}
\definecolor{blue}{HTML}{0563C1}
\definecolor{brightred}{HTML}{FF0000} 
\DeclareUnicodeCharacter{2212}{-}
\usepackage{hyperref,nameref}
\hypersetup{colorlinks=true,allcolors=blue}
\newcommand{\rref}[2]{\hyperref[#1]{\ref{#1}#2}} 

\color{black}
\usepackage[margin=0.67in]{geometry} 

\usepackage[switch]{lineno}   
\usepackage{amsmath,amssymb}  
\usepackage{textcomp} 
\usepackage{siunitx}
\usepackage{etoolbox} 
\newcommand*\linenomathpatch[1]{%
  \cspreto{#1}{\linenomath}%
  \cspreto{#1*}{\linenomath}%
  \csappto{end#1}{\endlinenomath}%
  \csappto{end#1*}{\endlinenomath}%
}
\linenomathpatch{equation}
\linenomathpatch{gather}
\linenomathpatch{multline}
\linenomathpatch{align}
\linenomathpatch{alignat}
\linenomathpatch{flalign}

\usepackage{fancyhdr,lastpage} 
\usepackage{titlesec}
\titleformat{\section}{\Large\bfseries}{\thesection}{0.5em}{}
\titleformat{\subsection}{\bfseries}{\thesubsection}{0.5em}{}
\titlespacing{\section}{0pt}{\baselineskip}{0pt}
\titlespacing*{\section}{0pt}{\baselineskip}{0pt}
\titlespacing*{\subsection}{0pt}{\baselineskip}{0pt}
\usepackage{float}

\usepackage[
  backend=biber,        
  autocite=superscript, 
  sorting=none,         
  style=numeric-comp,   
  maxnames=100,         
  minnames=100,         
  giveninits=false,     
  autopunct=false,      
  doi=true,             
  hyperref=true         
]{biblatex}
\DeclareFieldFormat{labelnumberwidth}{\;[#1]\!\!\!\!} 
\renewbibmacro{in:}{} 
\AtNextBibliography{\small} 
\AtEveryBibitem{\clearfield{pages}} 
\AtEveryBibitem{\clearfield{volume}} 
\AtEveryBibitem{\clearfield{number}} 
\AtEveryBibitem{\clearfield{month}} 
\AtEveryBibitem{\clearfield{eprint}} 
\AtEveryBibitem{\clearfield{issn}} 
\AtEveryBibitem{
    \clearfield{urlyear}
    \clearfield{urlmonth}
}

\usepackage[shortcuts, abbreviations,]{glossaries-extra}

\newabbreviation{SRO}{SRO}{short-range order}
\newabbreviation{SS}{SS}{solid solution}
\newabbreviation{fcc}{fcc}{face-centered cubic}
\newabbreviation{bcc}{bcc}{body-centered cubic}
\newabbreviation{hcp}{hcp}{hexagonal close-packed}
\newabbreviation{WC}{WC}{Warren-Cowley}
\newabbreviation{1CP}{1CP}{first coordination polyhedron}
\newabbreviation{KL}{KL}{Kullback-Leibler}
\newabbreviation{ML}{ML}{machine learning}
\newabbreviation{MC}{MC}{Monte Carlo}
\newabbreviation{1NN}{1NN}{first nearest neighbors}
\newabbreviation{MLP}{MLP}{machine learning potential}
\newabbreviation{DFT}{DFT}{density functional theory}

\glsdisablehyper 

\usepackage[]{cleveref} 
\creflabelformat{equation}{#2\textup{#1}#3} 

\usepackage{tabularx}
\usepackage{booktabs}
\newcolumntype{Y}{>{\raggedleft\arraybackslash}X} 
\usepackage{multirow}

\usepackage{mwe} 

\begin{document}

\twocolumn[
  \begin{center}
	\large
    \textbf{Bond-informed search for ordered configurations in multi-principal element alloys}
  \end{center}
  Gaurav Arora$^1${\footnotemark[1]},
  Alex Tai$^2$,
  Rodrigo Freitas$^2$,
  and David Cereceda$^1${\footnotemark[1]} \\
  $^1$\textit{\small Department of Mechanical Engineering, Villanova University, Villanova, PA, USA} \\
  $^2$\textit{\small Department of Materials Science and Engineering, Massachusetts Institute of Technology, Cambridge, MA, USA} \\
  
{\small Dated: \today}
\vspace{-0.15cm}
\begin{center}
\textbf{Abstract}
\end{center}
\vspace{-0.15cm}
Ordered configurations in multi-principal element alloys (MPEAs) control their local structure and influence a broad range of mechanical and thermodynamic properties. Yet they are notoriously difficult to identify.
Locating low-energy ordered candidates requires exploring a configurational space that grows combinatorially with the number of elements, and Monte Carlo searches driven by density-functional theory (DFT) need a new electronic-structure calculation at every trial step.
Here we introduce MECx (Minimum Energy Configuration eXplorer), a framework that guides Monte Carlo searches with a surrogate energy model trained on configurations stratified in bond-count space.
Applied to ternary, quaternary, and quinary BCC alloys using compact training sets of 100 DFT-labeled configurations, MECx identifies ordered configurations with lower energies than those found by direct DFT-based Monte Carlo within practical computational limits. These configurations exhibit distinct pair-specific ordering, including B2- and B32-like motifs, and undergo progressive disordering as temperature increases. The MPEA-trained models also transfer to selected binary subsystems without retraining. The resulting ordered configurations agree with those obtained by direct DFT-based Monte Carlo and produce experimentally distinguishable superlattice reflections in simulated neutron diffraction patterns.
\vspace{0.4cm}
] 
{
  \footnotetext[1]{Corresponding authors: (\texttt{gaurav.arora@villanova.edu, david.cereceda@villanova.edu}).}
}


\section*{Introduction}


Ordered configurations in multi-principal element alloys (MPEAs) emerge from a competition among a vast number of possible atomic arrangements on a common crystal lattice \cite{ye2016high, ma2021high}. Although these alloys are often described as random solid solutions, their equilibrium configurations can deviate systematically from a purely random distribution of elements, exhibiting short-range order (SRO), local clustering, and ordered motifs~\cite{george2019high,han2024multifunctional, santodonato2015deviation, zhang2020short, wang2023tailoring, chen2021direct}.
These departures from randomness influence stacking-fault and point-defect energies, tensile strength, work hardening, corrosion resistance, and irradiation-damage tolerance across a broad range of alloy systems~\cite{ding2018tunable,zhang2020short,li2019strengthening,xie2021percolation,zhang2023effect,chen2021direct,walsh2021magnetically,liu2018effect,scully2020controlling,blades2024tuning,arora2018effect}.
Uncovering the low-energy ordered configurations favored by a given alloy chemistry is therefore central to understanding and controlling the properties of MPEAs.

Yet such ordered configurations are notoriously difficult to identify.
Experimentally, their formation and observation depend on thermodynamic stability, transformation kinetics, processing history, and the sensitivity of the available characterization methods~\cite{han2024ubiquitous,stumpf2026nature,asle2026advances,zhang2020short,ferrari2023simulating,joress2023exafs,zhang2020short,walsh2023extra}.
Computationally, locating a candidate ordered phase requires searching for low-energy configurations on each relevant lattice within a configurational space that grows combinatorially with the number of elements.
%
Direct Monte Carlo (MC) searches driven by density-functional theory (DFT), referred to here as MC-DFT, provide a physically grounded route to this problem by evaluating the energy of each trial configuration with electronic-structure calculations during a Metropolis atom-swap search. However, this approach is computationally expensive: converging toward low-energy ordered configurations in 128-atom BCC supercells requires thousands of MC steps, each involving a full DFT evaluation~\cite{hastings1970monte,tamm2015atomic,ding2018tunable,zhang2020short,zhang2020robust, byggmastar2022simple,arora2025development,musa2025accelerating,jeffries2026generalized,kang2023density,zhang2017local,feng2017effects,ding2018tunable,zhang2020short,samanta2024machine,sheriff2024quantifying,cao2024capturing,yichen_mcdft_2024,qian2025hydrogen}. Accelerated MC-DFT approaches, such as the one proposed by Musa et al.~\cite{musa2025accelerating}, reduce this cost by replacing many DFT evaluations with surrogate-model predictions, but still require several hundred DFT-labeled configurations and couple the training data to the early stages of the MC trajectory, so that the surrogate inherits whatever region of configurational space the search happened to visit first.

Recent machine-learning approaches have highlighted the importance of descriptor design and data-efficient training strategies for extending first-principles predictions across compositionally complex alloy spaces~\cite{pathrudkar2025electronic,vazquez2024deciphering}. A more scalable route for configurational search is to construct the training set before the MC search begins, using a descriptor that captures the chemical environments most relevant to ordering. One such descriptor is the bond-count representation, defined by unordered elemental pair counts in the first and second nearest-neighbor coordination shells of the BCC lattice. On a fixed lattice, a linear model built on bond counts is equivalent to a pair cluster expansion truncated at the second nearest-neighbor shell~\cite{Kikuchi1951, Sanchez1984, Zunger2002, DeFontaine1994}, giving a physically interpretable and computationally compact representation of local chemical environments. Because the bond-count vector encodes the pair environments that drive ordering, it provides a natural space both for constructing training configurations and for training the energy surrogate model. How to distribute a small number of training configurations across this space is not obvious: random assignment of species to lattice sites concentrates near the disordered solid solution, while the ordered configurations of interest lie in the tails of the bond-count distribution.

Here we introduce the Minimum Energy Configuration eXplorer (MECx), a framework that guides finite-temperature Metropolis Monte Carlo searches with a ridge-regression surrogate trained on configurations sampled in bond-count space, to identify low-energy ordered configurations in MPEAs.  We compare random sampling~\cite{zhang2020robust}, motif-based sampling~\cite{sheriff2026machine}, and Latin hypercube sampling (LHS)~\cite{shields2016generalization, gramacy2024_spacefilling} for generating compact training sets, and find that stratifying the bond-count space with LHS gives the broadest coverage of ordered, clustered, and random-like configurations. Because the training set is constructed before the MC search begins, MECx is reproducible and requires no additional DFT evaluations during the MC search. We apply MECx to three BCC MPEAs of increasing chemical complexity: the ternary Cr-Mn-V alloy (MPEA3)~\cite{sultana2025investigation,arora2025development}, the quaternary Cr-Ta-Ti-W alloy (MPEA4)~\cite{sobieraj2020chemical}, and the quinary W-Ta-Cr-V-Hf alloy (MPEA5)~\cite{hea5-ref}. The results demonstrate that MECx achieves accurate DFT energy prediction from compact 100-structure training sets, reaches configurations with lower energies than those found by direct MC-DFT within practical computational limits, and is more data-efficient than accelerated MC-DFT~\cite{musa2025accelerating}. Warren-Cowley parameters computed along the MECx trajectories reveal pair-specific ordering tendencies and their temperature dependence from 0 to 1500~K, while simulated neutron diffraction patterns for selected binary subsystems show that the predicted ordering produces experimentally distinguishable superlattice signatures.

\section*{Results}
\subsection*{Bond-count representation of alloy configurational space}

\begin{figure*}[!b]
    \centering
    \includegraphics[width=0.85\textwidth]{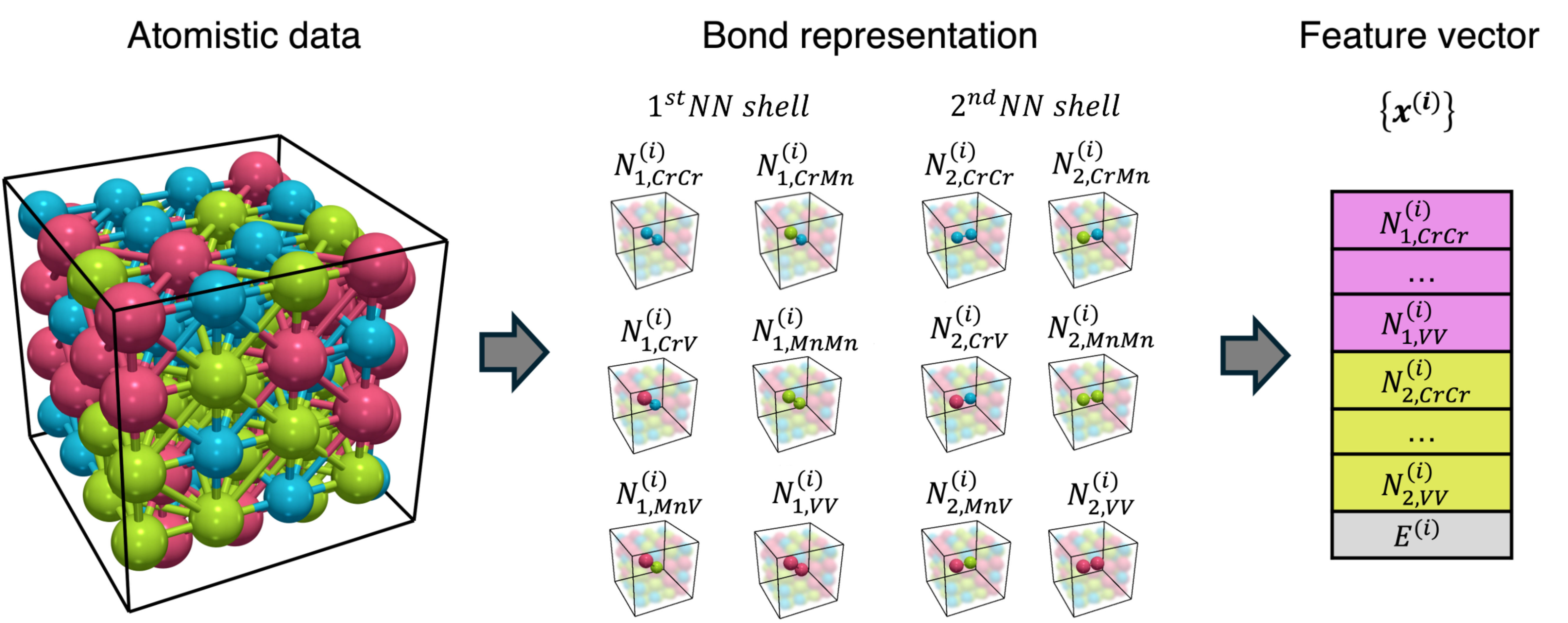}
    \caption{\textbf{Bond-count representation of atomic configurations in MECx.} An atomic configuration, shown here for Cr-Mn-V, is mapped onto a fixed-length descriptor by counting unordered $\alpha$--$\beta$ pairs in the first and second nearest-neighbor shells.
    For structure $i$, $N_{s,\alpha\beta}^{(i)}$ denotes the number of $\alpha$--$\beta$ bonds in coordination shell $s$.
    These shell-resolved bond counts are assembled into the feature vector $\mathbf{x}^{(i)}$, which is paired with the DFT total energy $E^{(i)}$ for surrogate-model training.}
    \label{fig:bond_descriptor}
\end{figure*}

We applied the framework to the three BCC alloys introduced above, MPEA3, MPEA4, and MPEA5. These systems were selected to test whether a descriptor-guided sampling and surrogate-modeling framework can handle alloys ranging from equiatomic ternary and quaternary compositions to a non-equiatomic refractory quinary alloy. All alloys were modeled in $4\times4\times4$ 128-atom supercells. MPEA3 and MPEA4 have nearly equiatomic compositions, whereas MPEA5 is non-equiatomic, with Cr and Hf present at lower concentrations than W, Ta, and V. Atomic percentages and exact supercell compositions are provided in Supplementary Table~S1.

To compare sampling strategies and train a compact energy surrogate on a common basis, each supercell was mapped onto a bond-count representation. In this representation, a configuration is described by the number of unordered elemental pairs in the first and second nearest-neighbor shells. This replaces the full site-occupation vector, whose length grows with the supercell, with a fixed-length vector whose entries are chemically interpretable. The descriptor is general to substitutional alloys on a fixed parent lattice, while the present work uses BCC MPEAs as representative case studies.

Figure~\ref{fig:bond_descriptor} illustrates the construction of the descriptor for a Cr-Mn-V supercell. The atomistic configuration is first converted into shell-resolved bond counts, where each component $N_{s,\alpha\beta}^{(i)}$ denotes the number of $\alpha$--$\beta$ bonds in coordination shell $s$ for structure $i$. The resulting counts are assembled into a fixed-length feature vector $\mathbf{x}^{(i)}$, which is paired with the DFT total energy $E^{(i)}$ and used as the input-output pair for surrogate-model training. Because the descriptor tracks chemically distinct nearest-neighbor environments, it provides a natural space to compare random sampling, motif-based sampling, and Latin hypercube sampling.

\subsection*{Comparing sampling strategies}

Having defined a common bond-count representation, we next compared random sampling (RS), motif-based sampling (MBS), and Latin hypercube sampling (LHS) in terms of the energy range and bond-count descriptor space sampled by each method. Random sampling assigns atomic species to lattice sites stochastically while preserving the prescribed composition. MBS, introduced by Sheriff et al.~\cite{sheriff2026machine}, aims to broaden the distribution of local chemical motifs. In contrast, LHS stratifies the bond-count descriptor space. This descriptor-space stratification allows LHS to access configurations farther from the disordered mean, including more ordered or clustered structures that are underrepresented by random sampling.

\begin{figure*}[!t]
    \centering
    \includegraphics[width=1\linewidth]{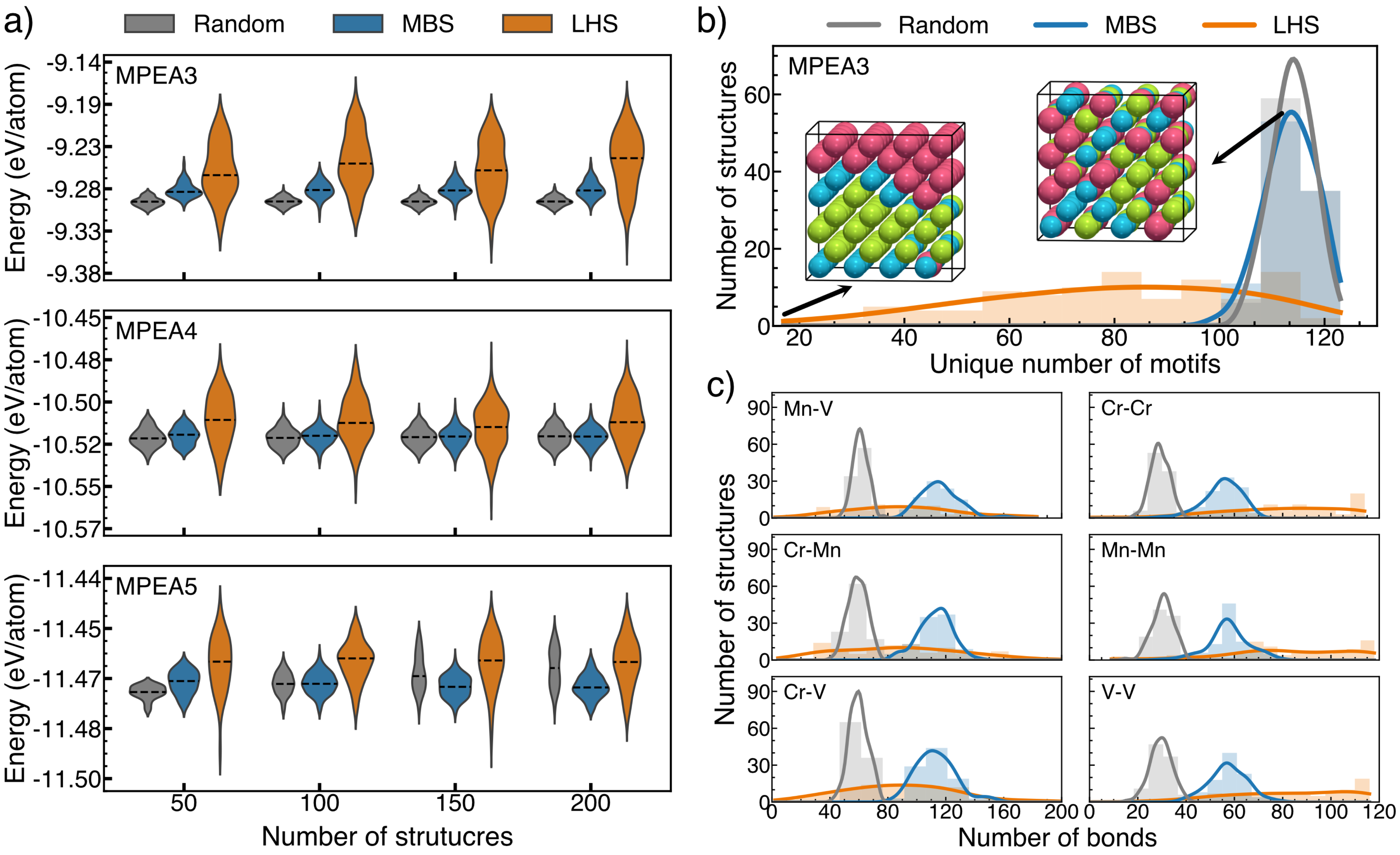}
    \caption{\textbf{Comparison of configurational distributions obtained with random sampling (RS), motif-based sampling (MBS), and Latin hypercube sampling (LHS).}
    (a) Violin plots showing the distribution of DFT total energies of sampled structures as a function of training-set size for MPEA3, MPEA4, and MPEA5. Dashed lines indicate the mean energy of that sample.
    Gray, blue, and orange distributions correspond to RS, MBS, and LHS, respectively.
    (b) Distribution of the number of unique chemical motifs for MPEA3 using 100 sampled structures. Insets show representative configurations with low and high motif diversity.
    (c) Bond-count distributions for the six unordered pair types in MPEA3 using 100 sampled structures.
    LHS produces the widest bond-count coverage, indicating more extensive exploration of ordered and clustered regions of configuration space.}
    \label{fig:sampling}
\end{figure*}

Figure~\ref{fig:sampling} compares the energy, motif, and bond-count coverage obtained with the three sampling strategies. For this sampling comparison, LHS structures were generated using only first nearest-neighbor (1NN) bond-count targets to match the 1NN motif definition used by MBS. Figure~\ref{fig:sampling}a shows the distribution of DFT total energies as a function of the number of sampled structures for MPEA3, MPEA4, and MPEA5. Across all three alloy systems and all dataset sizes considered, LHS produces the broadest energy distributions. For example, in MPEA3 with 100 structures, LHS spans an energy range of approximately 120 meV/atom, which is substantially larger than the corresponding ranges obtained with MBS and random sampling. This broader distribution indicates that stratifying the bond-count descriptor space enables LHS to sample configurations that are energetically more diverse than those obtained from methods concentrated near the random solid solution. The corresponding motif and bond-count comparisons for MPEA4 and MPEA5 are shown in Supplementary Figure~S1.

The difference between sampling strategies is also visible in the motif and bond-count distributions. Figure~\ref{fig:sampling}b shows the distribution of the number of unique chemical motifs in MPEA3 for 100 sampled structures. MBS broadens the range of motifs present in the dataset as a whole, with each individual MBS structure still containing many distinct motifs, as a disordered structure does. LHS, by contrast, spans a wider range of motif diversity across individual structures and includes configurations with far fewer distinct motifs, which is the signature of chemical ordering or clustering. The representative structures shown as insets illustrate this contrast: the low-motif LHS structure exhibits partial chemical ordering, whereas the high-motif MBS structure is more disordered.

Figure~\ref{fig:sampling}c shows the corresponding 1$^{\mathrm{st}}$NN bond-count distributions for the six unique pair types in MPEA3. LHS again provides the widest coverage, reaching both low and high bond-count tails for several pair types. This behavior follows from the construction of LHS in descriptor space: by enforcing stratified sampling along each bond-count dimension, LHS explores chemically ordered and clustered regions that are unlikely to be sampled efficiently by random assignment. Taken together, the energy, motif, and bond-count distributions show that LHS provides broader coverage of alloy configurational space than both RS and MBS. We therefore used LHS-generated configurations as the training data for the surrogate model developed below. For this purpose, LHS was applied in the full descriptor space spanning both the 1$^{\mathrm{st}}$NN and second nearest-neighbor (2$^{\mathrm{nd}}$NN) shells, rather than the 1$^{\mathrm{st}}$NN-only space used for the comparison above.

\subsection*{MECx energy predictions and DFT validation}

The broader energy and bond-count coverage obtained with LHS suggests that accurate energy models can be trained from relatively compact DFT datasets. To test this, we trained a ridge-regression surrogate on LHS-generated configurations, with the bond-count descriptor defined above as input and the DFT total energy as the target. Together with the Monte Carlo configurational search described below, the descriptor-guided sampling strategy and surrogate model constitute the Minimum Energy Configuration eXplorer (MECx) framework.

\begin{figure*}[!t]
    \centering
    \includegraphics[width=0.90\linewidth]{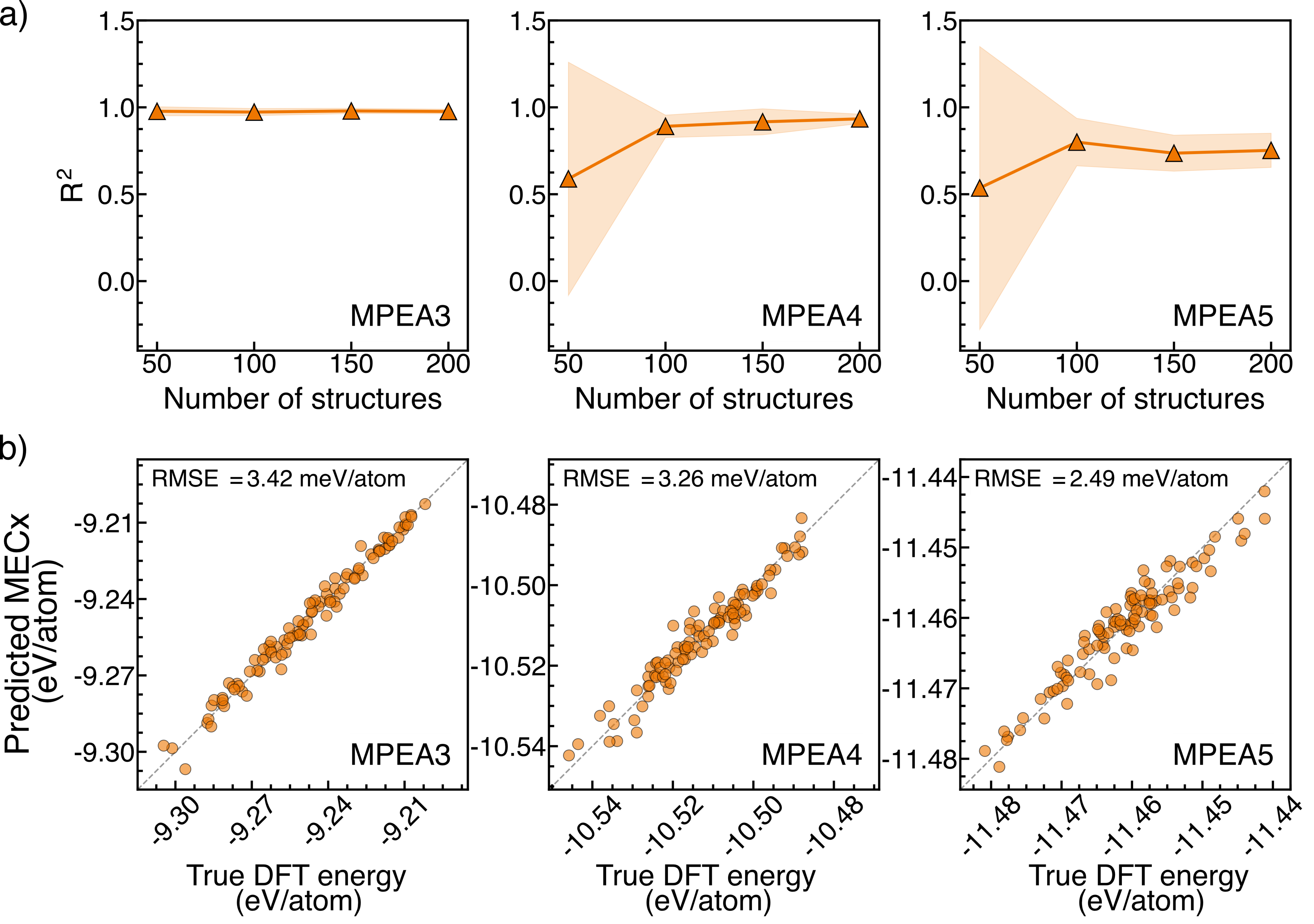}
    \caption{\textbf{Accuracy of the ridge-regression surrogate model.} (a) Coefficient of determination $R^2$ of the ridge-regression surrogate as a function of the number of training structures for MPEA3, MPEA4, and MPEA5. Solid lines show the mean and shaded bands show the standard deviation from five independent repetitions of 10-fold cross-validation. (b) Parity plots comparing the MECx predicted energies against the DFT energies for the three systems, evaluated with a training set of 100 structures. The dashed line marks perfect agreement, and the root mean square error (RMSE) is reported in each panel.}
    \label{fig3}
\end{figure*}

We first determined the number of LHS structures required to train an accurate energy model. Figure~\ref{fig3}a shows the validation $R^2$ score as a function of training-set size for the three alloy systems. For MPEA4 and MPEA5, the $R^2$ score increases sharply as the number of training structures increases from 50 to 100, while the fold-to-fold variation decreases. Increasing the training set beyond 100 structures produces only modest additional improvement. For MPEA3, the validation accuracy is already high with 50 structures and remains nearly unchanged up to 200 structures, consistent with the smaller number of coefficients to be fitted (12 bond-count features for MPEA3 against 20 and 30 for MPEA4 and MPEA5).

Figure~\ref{fig3}b shows parity plots comparing MECx-predicted energies with DFT energies for models trained using 100 LHS structures. The predicted and DFT energies show close agreement across all three systems. The largest root mean square error (RMSE) is 3.42 meV/atom for MPEA3, and the lowest $R^2$ value is 0.935 for MPEA4. These results indicate that the bond-count surrogate captures the dominant DFT energy trends using only 100 training structures.

\subsection*{Identification of minimum-energy configurations}
Having established that MECx accurately reproduces DFT energy trends from compact LHS training sets, we next used the trained surrogate models to search for low-energy configurations in each alloy system. The search was performed by Metropolis Monte Carlo atom swaps at fixed composition, using the ridge-regression surrogate to evaluate the energy of each trial configuration. We benchmarked this surrogate-driven search against direct MC-DFT, in which each trial configuration is evaluated with a full DFT calculation.

\begin{figure*}[!t]
    \centering
    \includegraphics[width=1\linewidth]{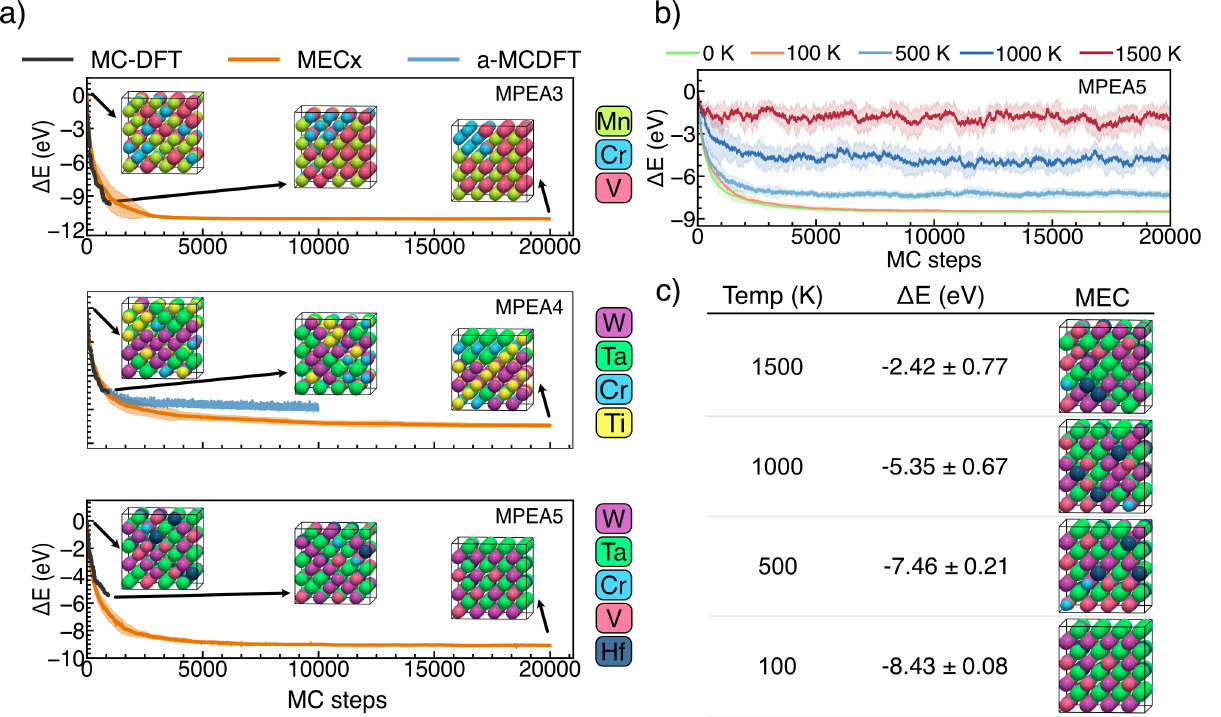}
    \caption{\textbf{Energy evolution and temperature dependence during Monte Carlo configurational search}. (a) Energy change $\Delta E$ relative to the initial configuration as a function of Monte Carlo steps at 100\,K for MPEA3, MPEA4, and MPEA5. Results from MC-DFT (black), MECx (orange), and a-MCDFT (light blue) are compared, where a-MCDFT data are available only for MPEA4. The shaded bands show the standard deviation over ten independent MECx runs. For each system, the left and right insets show the initial and final MECx configurations, and the middle inset shows the final MC-DFT configuration after 1000 steps. The element color keys are given on the right. (b) $\Delta E$ as a function of Monte Carlo steps for MPEA5 at 0, 100, 500, 1000, and 1500\,K. Solid lines show the mean and shaded bands show the standard deviation over ten independent runs. (c) Final $\Delta E$ for MPEA5 at 1500, 1000, 500, and 100\,K, together with the corresponding minimum-energy configurations. Values are reported as mean $\pm$ standard deviation.}
    \label{fig4}
\end{figure*}

Figure~\ref{fig4}a compares the energy evolution obtained with MECx and MC-DFT at 100\,K for MPEA3, MPEA4, and MPEA5. Direct MC-DFT is computationally expensive and was therefore limited to 1000 Monte Carlo steps, whereas MECx was run for 20000 steps at negligible additional cost after training. During the first 1000 steps, MECx and MC-DFT follow similar energy-lowering trajectories, indicating that the surrogate captures the local energetic driving forces sampled by the direct DFT search. Beyond this point, MECx continues to explore the configurational landscape and reaches substantially lower energies than those accessible within the MC-DFT step budget.

For MPEA4, we also compared MECx with the accelerated Monte Carlo DFT approach a-MCDFT~\cite{musa2025accelerating}, for which literature data is available. MECx reaches lower energies than a-MCDFT while requiring only 100 DFT-labeled training structures, compared with 300 structures used in a-MCDFT. This improvement likely follows from the broader descriptor-space coverage provided by LHS, which allows the surrogate to be trained from a smaller but more diverse set of configurations. The supercell snapshots in Fig.~\ref{fig4}a further show that the final MECx configurations develop clear chemical ordering relative to the initial random-like structures.

To validate the configurations identified by MECx, we performed DFT calculations on both the initial SQS and the lowest-energy configuration obtained across the MECx runs. The latter configuration is referred to as the MEC. Across the three alloy systems, the absolute difference between MECx-predicted and DFT-computed energies ranged from 2.8 to 17.8 meV/atom, which corresponds to 6\%--37\% of the ordering energy of each alloy. The full benchmark comparison is provided in Supplementary Table~S2. This agreement shows that MECx remains sufficiently accurate to guide the search toward low-energy configurations not explicitly included in the training set.

In addition to accurately predicting energies and identifying low-energy configurations, MECx substantially reduces the computational cost of the Monte Carlo search. In MC-DFT, the cost increases linearly with the number of Monte Carlo steps because each trial configuration requires a new DFT calculation, which extrapolates to approximately $10^{4}$ CPU hours on 256 cores at 20000 steps. In contrast, MECx incurs most of its cost during construction of the fixed DFT training set and subsequent Monte Carlo steps are evaluated by the surrogate model at a cost more than two orders of magnitude below that of MC-DFT at 20000 steps (Supplementary Figure~S3). MECx requires approximately three times fewer DFT-labeled training structures than a-MCDFT \cite{musa2025accelerating}. 

We then used this efficiency to examine how temperature affects both the energy reduction and the degree of chemical ordering during the MECx search. Figure~\ref{fig4}b shows the energy change for MPEA5 over 20{,}000 Monte Carlo steps at temperatures from 0 to 1500\,K. At low temperature, the energy decreases rapidly and converges to a substantially lower value, indicating that the system moves away from the initial configuration toward a chemically ordered low-energy configuration. At 1500\,K, by contrast, the energy decrease is much smaller, because at this temperature the thermodynamically stable state is a solid solution with chemical short-range order, whose configurational entropy outweighs the lower energy of the intermetallic phase.

Figure~\ref{fig4}c summarizes this temperature dependence by showing the final energy change ($\Delta E$) together with the corresponding minimum-energy configurations. The magnitude of the energy reduction increases systematically as temperature decreases, from $-2.4 \pm 0.8$ eV at 1500\,K to $-8.4 \pm 0.1$ eV at 100\,K. The associated configurations reveal a parallel structural trend: the 1500\,K configuration remains largely chemically disordered, whereas the 100\,K configuration exhibits pronounced chemical ordering. The run-to-run variability also decreases at lower temperature, suggesting more consistent convergence toward similar ordered low-energy states. Simulations at 0\,K gave nearly identical results to those at 100\,K and are shown in Fig.~\ref{fig4}b for completeness. Temperature-dependent results for MPEA3 and MPEA4 are provided in Supplementary Figure~S2. 
Taken together, these results show that MECx efficiently explores complex alloy configurational landscapes, identifying lower-energy ordered configurations that neither MC-DFT nor a-MCDFT reached within comparable computational budgets.

\subsection*{Ordered configurations and their thermal disordering}

\begin{figure*}[!t]
    \centering
    \includegraphics[width=1\linewidth]{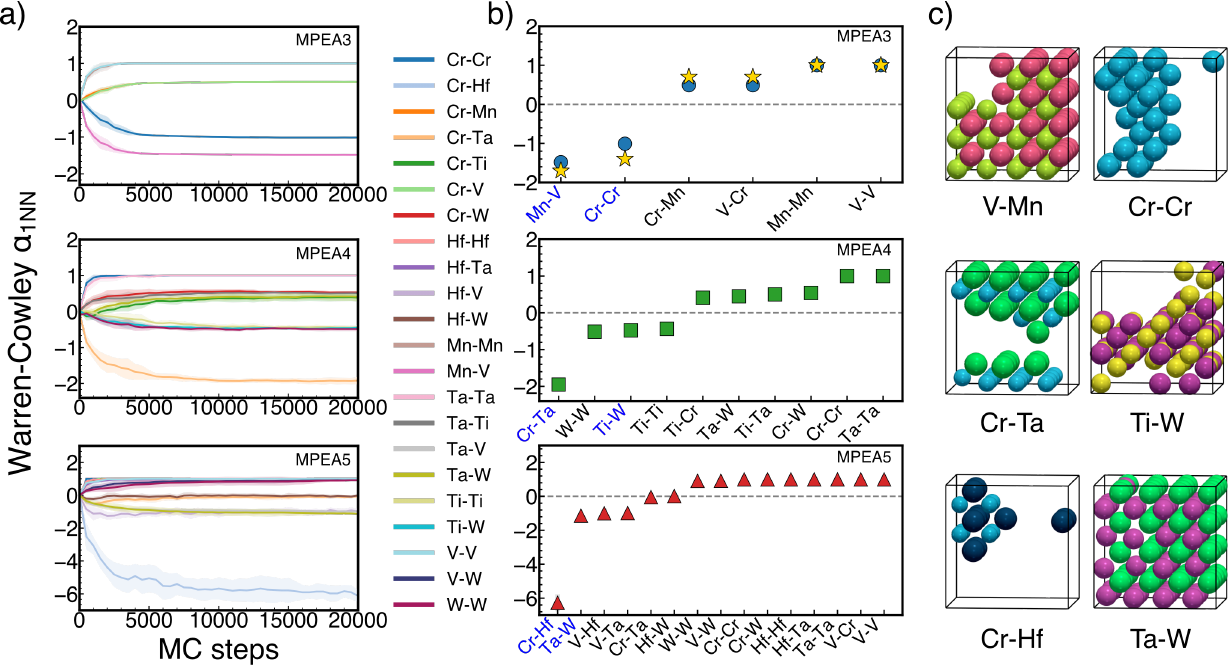}
    \caption{\textbf{Pair-specific ordering in configurations identified by MECx.}
    (a) Evolution of the 1st NN shell Warren-Cowley parameter $\alpha_{1\mathrm{NN}}$ as a function of Monte Carlo steps for all atomic pairs in MPEA3, MPEA4, and MPEA5. Solid lines show the mean and shaded bands show the standard deviation over ten independent 
    runs. (b) Converged $\alpha_{1\mathrm{NN}}$ values for each pair, sorted in ascending order for the three alloys. For MPEA3, circles show values from this work and stars show values from a machine-learned interatomic potential~\cite{arora2025development}. (c) Converged configurations at 100 K with the sublattices formed by the Mn-V, Cr-Ta, and Ta-W pairs (B2-like), the Ti-W pair (B32-like), the Cr-Hf pair, and the Cr-Cr clusters highlighted.}
    \label{fig5}
\end{figure*}

The low-energy configurations identified by MECx suggest that the energy reduction observed during the Monte Carlo search is accompanied by chemical ordering, and now we seek to identify the phases found by this search. Chemical order in a substitutional alloy comes in two degrees. In an intermetallic phase the species occupy distinct sublattices, so the order is long-ranged. In a solid solution there are no sublattices, but neighboring sites can still be correlated, with unlike (or like) pairs more frequent than in a random arrangement (i.e., chemical short-range order). The Warren-Cowley parameter of a pair measures the deviation of its bond count from a random arrangement: it is zero for a random arrangement, negative when the two species attract (the pair appears more often than at random), and positive when they repel, reaching its bound only for full long-range order (see Methods). We computed it for each atomic pair along the MECx trajectories in the 1NN and 2NN shells ($\alpha_{\mathrm{1NN}}$ and $\alpha_{\mathrm{2NN}}$) and as their coordination-weighted average ($\alpha_{\mathrm{avg}}$). Note that these parameters are a normalization of the bond counts that form the MECx descriptor, so they are the observable a pair cluster expansion controls most directly. Figure~\ref{fig5}a shows the evolution of $\alpha_{1\mathrm{NN}}$ as a function of Monte Carlo steps for MPEA3, MPEA4, and MPEA5. All pair parameters start near zero, as expected for the random initial configurations, and evolve toward steady-state equilibrium values with 5{,}000--10{,}000 steps. This convergence occurs well within the 20{,}000-step trajectories, and the converged parameters characterize the phase reached by the search, which we examine next.

Figure~\ref{fig5}c shows the converged configurations, with the sublattices formed by selected pairs highlighted. The Mn-V, Cr-Ta, and W-Ta pairs form B2-like sublattices, in which each atom is surrounded by unlike 1NN neighbors, while W-Ti forms a B32-like arrangement, in which the 2NN shell orders as well. The Cr-Hf pair also orders strongly, but Cr and Hf are dilute in MPEA5 and do not form a sublattice that fills the cell, so no crystallographic assignment is possible. In contrast, Cr-Cr shows clustering behavior, consistent with its negative Warren-Cowley parameter for like atoms. These assignments are qualitative: in a five-element 128-atom cell several ordered arrangements are consistent with the same pair correlations, which is why
the binary systems are examined independently through simulated neutron diffraction in the following subsection. The corresponding second-nearest-neighbor and shell-averaged Warren-Cowley parameters are provided in Supplementary Figure~S4 and S5. The averaged values preserve the same dominant ordering trends.

Which pairs drive each phase is read from Fig.~\ref{fig5}b, which ranks the converged $\alpha_{1\mathrm{NN}}$ values of all pairs in each alloy. The pair with the most negative $\alpha_{1\mathrm{NN}}$, i.e. the one that most strongly seeks unlike neighbors, differs across the three alloys: Mn-V in MPEA3, Cr-Ta in MPEA4, and Cr-Hf in MPEA5. The Cr-Hf pair has the most negative value of any pair in any of the three alloys, in agreement with prior reports that identify Cr-Hf as the dominant ordering pair in this system~\cite{hea5-ref}. For MPEA3, the Warren-Cowley parameters obtained with MECx agree closely with literature values~\cite{arora2025development}, which is an independent check of the surrogate against a machine-learned interatomic potential.

\begin{figure*}[!t]
    \centering
    \includegraphics[width=1\linewidth]{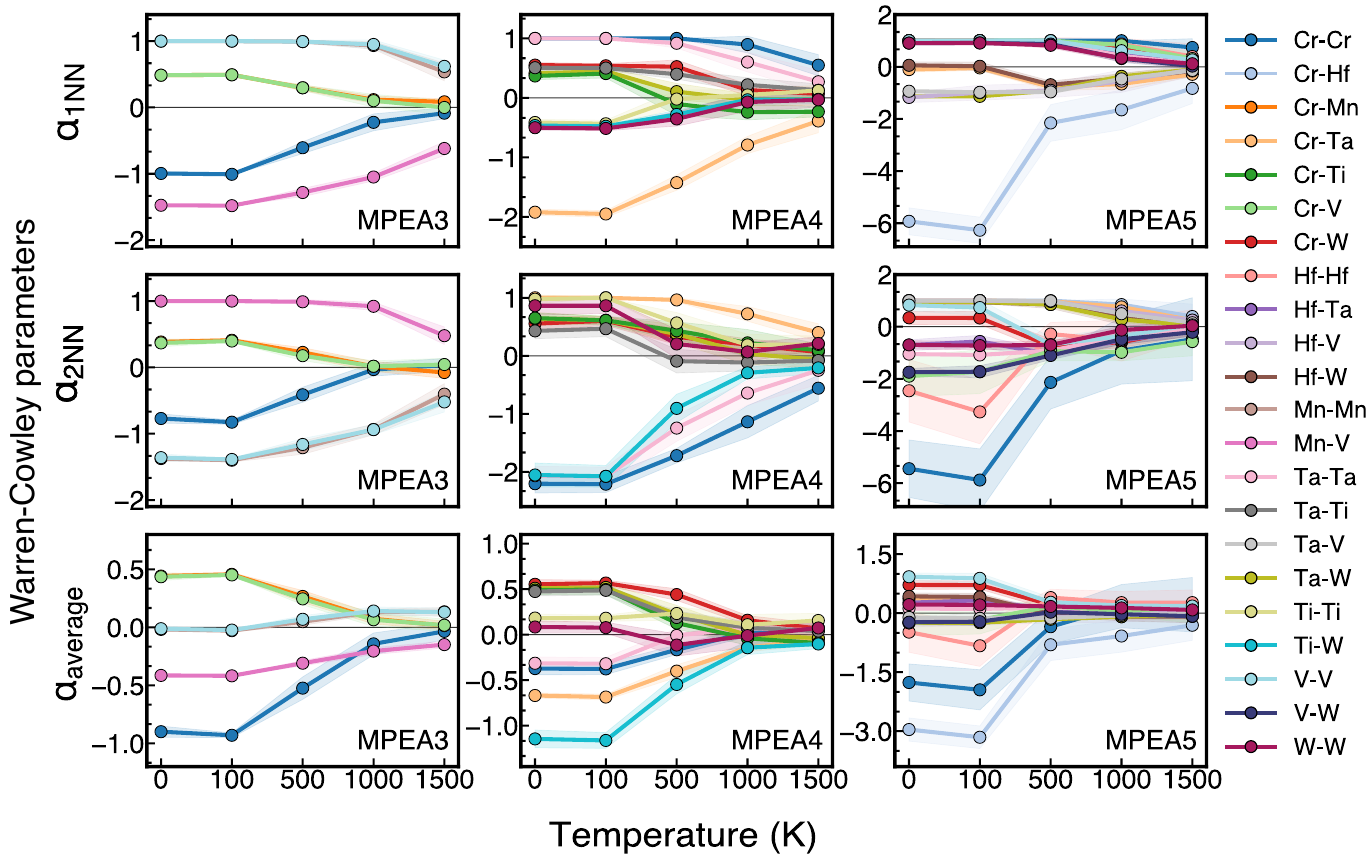}
    \caption{Temperature dependence of Warren-Cowley parameters for MPEA3, MPEA4, and MPEA5.
    The top row shows $\alpha_{\mathrm{1NN}}$, the middle row shows $\alpha_{\mathrm{2NN}}$, and the bottom row shows $\alpha_{\mathrm{avg}}$, the coordination-weighted average over the 1st and 2nd NN shells. Each curve corresponds to one atomic pair. Markers show mean values 
    at 0, 100, 500, 1000, and 1500\,K, and shaded bands show the standard deviation over ten independent runs. Increasing temperature drives the parameters toward zero, marking the transition from the ordered phase to a solid solution with short-range order.}
    \label{fig6}
\end{figure*}

We next examined how the ordered phases give way to the solid solution as temperature increases. Figure~\ref{fig6} shows the temperature dependence of $\alpha_{\mathrm{1NN}}$, $\alpha_{\mathrm{2NN}}$, and $\alpha_{\mathrm{avg}}$ for all three alloy systems. At low temperature, several pair parameters have large magnitudes, as expected for the sublattice occupation of the ordered phases identified above. As temperature increases, the Warren-Cowley parameters move progressively toward zero, which is the passage from the intermetallic phase to a solid solution that retains SRO but no sublattices. This is the same competition seen in the energy trajectories of Figure~\ref{fig4}b,c: the configurational entropy of the solid solution outweighs the lower energy of the ordered phase once the temperature is high enough. A fixed-temperature Metropolis search in a 128-atom cell places this transition only approximately, so we draw conclusions from the ordering trend with temperature rather than from the transition temperature itself.

\subsection*{Diffraction signatures of the predicted phases}

\begin{figure*}[!t]
    \centering
    \includegraphics[width=0.95\linewidth]{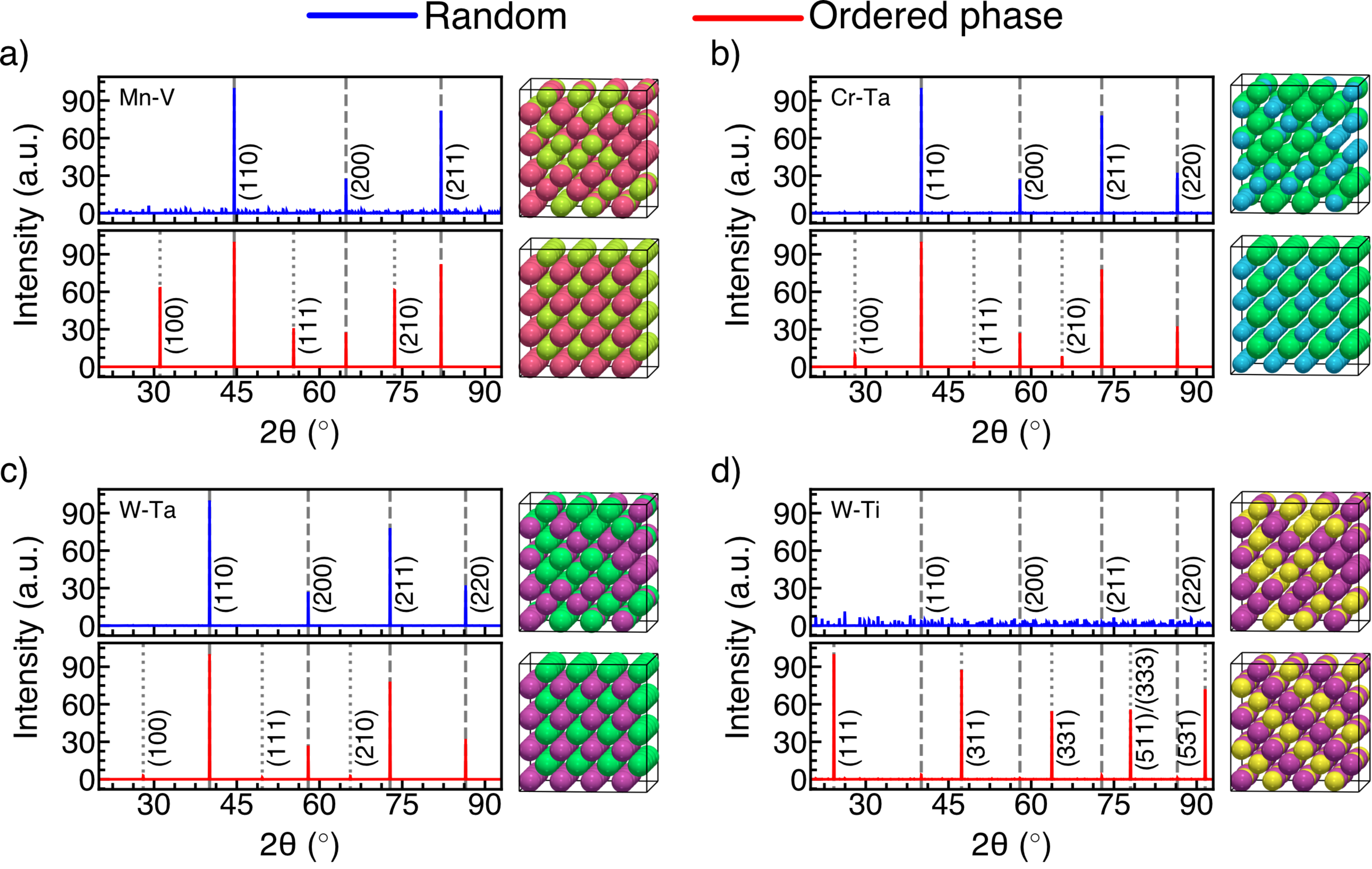}
    \caption{Simulated neutron diffraction patterns for the binary alloys (a) Mn-V, (b) Cr-Ta, (c) W-Ta, and (d) W-Ti (see Methods for simulation details). In each panel, the upper pattern (blue) corresponds to the random solid solution and the lower pattern (red) corresponds to the ordered phase found by MECx. Dashed lines mark the fundamental BCC reflections common to both structures, and dotted lines mark the additional superlattice reflections of the ordered phase. Peaks are indexed by their Miller indices. Insets show the corresponding random (top) and ordered (bottom) configurations.}
    \label{fig7}
\end{figure*}

An intermetallic phase is confirmed experimentally by diffraction, through superlattice reflections that the solid solution lacks. We next asked whether the phases predicted above would produce such reflections~\cite{zaeem2026advances}.
We did this in the equiatomic binaries of the strongly ordering pairs, where the crystallographic assignment is unambiguous and where the corresponding intermetallic is known or expected to form: Mn-V, Cr-Ta, W-Ta, and W-Ti. Although Cr-Hf exhibited a strong ordering tendency in MPEA5, we did not include it because Cr and Hf are dilute in MPEA5, so an equiatomic Cr-Hf binary would be a larger compositional extrapolation than the other binaries. No retraining is needed: the surrogate carries a coefficient for every pair, so it can be evaluated at any composition of the same elements. 

Because the surrogate is extrapolated in composition, we first checked the binary phases it predicts against MC-DFT searches of up to 1{,}000 steps for the same four systems. The DFT search converges toward the same ordered phases as MECx: B2 for Mn-V, Cr-Ta, and W-Ta, and B32 for W-Ti as shown in Figure~S6. The surrogate therefore captures the energetic driving force for ordering in these pairs, not a numerical artifact of the fit.

For each selected binary, we computed simulated neutron diffraction patterns for the random solid solution and the ordered phase found by MECx. Neutrons were chosen because their coherent scattering lengths give chemical contrast for these pairs, where X-rays give little contrast. Figure~\ref{fig7} compares the simulated diffraction patterns for Mn-V, Cr-Ta, W-Ta, and W-Ti. In each panel,  the random solid solution shows only the fundamental BCC reflections, marked by dashed lines, whereas the ordered configuration exhibits additional superlattice reflections, marked by dotted lines, at positions absent from the chemically disordered BCC structure. For the B2-like Mn-V, Cr-Ta, and W-Ta configurations, these include the 100, 111, and 210 reflections. For the B32-like W-Ti configuration, which has a doubled cubic unit cell, the superlattice reflections are indexed as 111, 311, 331, 511/333, and 531 in the B32 cell. The W-Ti binary provides a particularly clear example: the random pattern is nearly flat because the coherent neutron scattering lengths of Ti and W largely cancel for the fundamental reflections, while the ordered phase produces strong superlattice peaks. Thus, if realized experimentally, the MECx-identified configurations would be distinguishable from their corresponding disordered BCC solid solutions through the appearance of superlattice reflections.

\section*{Discussion}

MECx finds ordered configurations of each alloy from 100 DFT calculations, at energies that MC-DFT does not reach within its step budget. Why a linear model of pair counts trained on so few structures suffices warrants further explanation. On a fixed BCC lattice, the bond-count descriptor is equivalent to a pair cluster expansion truncated at the second nearest-neighbor shell, so the energy it assigns is additive in the bond counts. A model of this form can rank configurations it has never seen, including ones far more ordered than any in its training set, provided its pair coefficients are right~\cite{cowley1950approximate,clapp1971pair}. The pair coefficients are determined by how the training set is distributed. Random sampling concentrates near the solid solution, where all configurations have nearly the same bond counts and the fit is poorly conditioned. LHS instead places training points in the ordered and clustered tails of the bond-count space, where the differences between pair energies are large and the coefficients are constrained.

The clearest evidence that the model extrapolates is that the phases it finds lie below the energy range of its training set. For MPEA5, the ordered phase identified by MECx has a DFT-computed energy approximately 36 meV/atom below the lowest-energy structure in the training set, corresponding to a prediction error of 17.8 meV/atom, the largest among the six benchmark cases in Supplementary Table~S2 and consistent with the surrogate operating outside its training distribution. This error is about one-third of the 46.8 meV/atom ordering energy of MPEA5, small enough for the surrogate to rank the ordered phase below every configuration it competes with, but not small enough for its absolute energy to be trusted. For this reason the DFT recomputation of the final configuration is part of the MECx protocol.

The same argument explains the comparison with a-MCDFT~\cite{musa2025accelerating}, which reached higher energies for MPEA4 with three times as many DFT-labeled structures. a-MCDFT builds its training set from configurations visited early in the search, which by construction lie near the solid solution, so its surrogate sees the ordered tail only as the search reaches it. Constructing the training set before the search also makes MECx reproducible: two runs with the same training set search the same energy model, whereas two a-MCDFT runs that diverge early are fitting different models.

The ordered configurations found by MECx can be assessed against two references: prior DFT studies constrained to the BCC lattice and experimental binary phase diagrams.
The B2-like ordering of Mn-V, Cr-Ta, and W-Ta and the B32-like ordering of W-Ti agree with prior DFT studies of ordering on the fixed BCC lattice~\cite{sobieraj2020chemical,alvarado2023predicting,arora2025development}, which, sharing the same lattice constraint, provide the appropriate benchmark for the present calculations. 
Experimental binary phase diagrams provide a broader test because they include competing phases and crystal structures outside the present BCC search space~\cite{massalski1990binary}. 
For Mn-V, a B2 ordered phase and its order--disorder relationship with the BCC solid solution have been reported, consistent with the B2-like configuration identified by MECx. For Cr-Ta, the reported intermediate compounds are TaCr$_2$ Laves phases rather than an equiatomic B2 phase. W-Ta is reported primarily as a BCC solid-solution system without a stable equiatomic intermetallic compound, whereas the Ti-W phase diagram contains BCC and Ti-rich HCP phase fields rather than a reported equilibrium B32 compound. The Cr-Ta, W-Ta, and W-Ti configurations identified here should therefore be interpreted as preferred ordering motifs within the fixed BCC lattice, rather than as predictions of globally stable binary phases.
Nothing in MECx is specific to BCC, however. The bond-count descriptor is defined for any parent lattice, so the same protocol run on the competing lattices of a given system would yield the candidate ordered phases of each, and comparing them is a matter of DFT energies of the final configurations. The present work demonstrates the protocol on one lattice. Within that lattice, the temperature dependence of the Warren-Cowley parameters traces the order-disorder transition from each intermetallic phase to a solid solution with SRO, driven by the configurational entropy of the latter. 

A further result of this work is that the MPEA-trained surrogate transfers to binary subsystems composed of the same elements. Because the bond-count descriptor is defined over elemental pair counts rather than full multicomponent configurations, the surrogate trained on a parent MPEA can be evaluated at any composition of the same elements, without retraining. For the four binaries examined here, the ordered configurations found this way agree with MC-DFT and produce the superlattice reflections by which they would be identified in a diffraction experiment. A single training set of 100 DFT calculations on the parent alloy thus screens all of its binary subsystems for BCC-derived intermetallics, and the same approach should extend to its ternary and quaternary subsystems, which were not examined here.

Two approximations underlie the present approach. The first is the pair truncation. A bond-count model contains no term for many-body or angular correlations, so an alloy whose ordering is driven by them, or by pronounced lattice relaxation, requires a richer descriptor. While the pair model reaches DFT accuracy of a few meV/atom in the three alloys studied here, which indicates that pair interactions dominate their ordering, the same need not hold for other systems. The second is the electronic-structure setup: DFT calculations were performed without spin polarization and with a single $\Gamma$-centered $k$-point mesh to maintain high-throughput efficiency for training-set generation, which limits the absolute accuracy of the reference energies and may affect the predicted ordering tendencies for magnetically active systems. The framework also neglects vacancy formation, interstitial defects, and composition-dependent lattice parameter changes~\cite{nataraj2021systematic}. 

Taken together, MECx finds the intermetallic phases of a multicomponent alloy, and of its subsystems, from compact DFT datasets. Applied lattice by lattice to the structures that compete in a given system, it gives the candidate ordered phases whose energies a phase diagram is built from. Future work could incorporate lattice relaxation effects, add three-body terms to the descriptor where pair interactions prove insufficient, or combine the bond-count surrogate with interatomic potentials to enable size-converged predictions of ordering length scales and their effects on mechanical properties. More broadly, descriptor-guided configurational search provides a practical path toward connecting alloy chemistry, local ordering tendencies, and experimentally observable structural signatures in chemically complex materials.

\printbibliography[heading=bibintoc,title={References}]

\end{document}